# Observation of Chiral State Transfer Without Encircling an Exceptional Point

Hadiseh Nasari[1,2], Gisela Lopez-Galmiche[2], Helena E. Lopez-Aviles[2], Alexander Schumer[1,3], Absar U. Hassan[2], Qi Zhong[2], Stefan Rotter[3], Patrick LiKamWa[2], Demetrios N. Christodoulides[2], Mercedeh Khajavikhan[1,4,*]

[1]Ming Hsieh Department of Electrical and Computer Engineering, University of Southern California, Los Angeles, California 90089, USA.
[2]CREOL, The College of Optics & Photonics, University of Central Florida, Orlando, Florida 32816–2700, USA.
[3]Institute for Theoretical Physics, Vienna University of Technology (TU Wien), Vienna, A-1040, Austria.
[4]Department of Physics & Astronomy, University of Southern California, Los Angeles, California 90089, USA.

*Corresponding author: khajavik@usc.edu (M.K.)

**The adiabatic theorem, a corollary of the Schrödinger equation, manifests itself in a profoundly different way in non-Hermitian arrangements, resulting in counterintuitive state transfer schemes that have no counterpart in closed quantum systems. In particular, the dynamical encirclement of exceptional points (EPs) in parameter space has been shown to lead to a chiral phase accumulation, non-adiabatic jumps, and topological mode conversion[1-8]. Recent theoretical studies, however, have shown that contrary to previously established demonstrations, this behavior is not strictly a result of winding around a non-Hermitian degeneracy[9]. Instead, it appears to be mostly attributed to the non-trivial landscape of the Riemann surfaces, sometimes because of the presence of an exceptional point in the vicinity[9-11]. In an effort to bring this counterintuitive aspect of non-Hermitian systems into light and confirm this hypothesis, we provide here the first set of experiments to directly observe the field evolution and chiral state conversion in an EP-excluding cycle in a slowly varying non-Hermitian system. To do so, a versatile yet unique fiber-based photonic emulator is realized that utilizes the polarization degrees of freedom in a quasi-common path single-ring arrangement. Our observations may open up new avenues for light manipulation and state conversion, while providing a foundation for understanding the intricacies of the adiabatic theorem in non-Hermitian systems.**

The energy Riemann surfaces associated with non-Hermitian Hamiltonians exhibit a complex non-trivial topology that is strongly affected by the presence of non-Hermitian degeneracies [12-19]. This is shown to enable non-trivial state transfer dynamics in slowly evolving non-Hermitian systems when an exceptional point (EP) is encircled[1-11, 20-24]. For example, experimental works in two-state exciton-polariton systems[25] and in microwave cavities[26] show that winding around an EP under quasi-static conditions leads to a mode exchange occurring at the end of a parametric loop, accompanied by an effective geometric phase of $\pi$ for one of the modes[27-29]. For a fully dynamical loop around an EP, the contribution of non-Hermiticity and its inevitable concomitant, namely non-adiabatic transitions, give rise to a fundamentally different chiral behavior. In this case, if the encirclement is carried out in a clockwise (CW) fashion, for any arbitrary input that happens to be a mixture of the two eigenstates, the system always ends in one of the eigenmodes. On the other hand, if the loop is traversed in the opposite direction, i.e., in a counterclockwise (CCW) manner, the second eigenstate is robustly produced, irrespective of the input state composition[1-11, 20-24]. Analytical studies using confluent hypergeometric functions provide a way to justify this striking behavior[7, 9]. In the past few years, dynamical EP encirclement and the ensuing mode conversion were reported in lossy microwave waveguide arrangements with a judiciously patterned boundary modulation[4, 11], in cryogenic optomechanical oscillators[5], as well as in silicon photonic based optical waveguides[6]. Finally, while a remarkable progress has been made in monitoring the field intensities of radio frequency signals during the encirclement process[24] a direct observation of the evolution of the full state of light in these dynamical systems is still lacking.

Recent theoretical studies, however, have challenged these findings by showing that the counterintuitive chiral state transfer in these systems is not necessarily associated with encircling an exceptional point[9-11]. Instead, the topology and landscape of the intersecting Riemann surfaces and the trajectory and speed of the steering process greatly influence the outcome of a parametric loop. For example, without strictly winding an EP, one may still be able to observe a chiral behavior, or dynamically encircling an EP along two dissimilar paths in the parameter space may result in different outputs, when multiple exceptional points are involved[9-11, 30]. In this Letter, we report on directly probing the temporal evolution of light in systems with slowly varying non-Hermitian Hamiltonians, in order to render a more complete portrait of the physics of quasi-adiabatic steering. In particular, we monitor the polarization state of a laser pulse on the Poincaré sphere as it circulates in a time-varying fiber ring. By imposing intensity and phase modulation as well as polarization mixing, each round trip in the ring effectively represents a new point in the parameter space. Consequently, a judicious modulation pattern over consecutive round trips allows the system to emulate an arbitrary parametric loop. We note that this approach is different from that used with standard spatially non-uniform coupled waveguide systems where information about the state of the light is only available at the input and output ports. Instead, through this unique arrangement, we are able to observe, for the first time, the light's full state evolution as it undergoes adiabatic transformations and non-adiabatic jumps in the vicinity of an exceptional point without encircling it.

In our study, the dynamics of the time-varying light propagation process is governed by a discrete analogue of the Schrödinger equation $|\psi_{n+1}\rangle = (1 - iH_n)|\psi_n\rangle$ where $|\psi_n\rangle = [a_n, b_n]^T$ is the Jones vector representation of the polarization state with field amplitudes $a_n$ and $b_n$ at the $n$th round trip (Supplementary Section 1). The associated binary non-Hermitian Hamiltonian corresponding to the $n$th circulation in the fiber is given by

$$H_n = \begin{pmatrix} ig_n & \kappa \\ \kappa & \delta_n \end{pmatrix}, \tag{1}$$

where $\kappa$ denotes a fixed coupling strength between the two orthogonal polarization constituents. Unlike in strict parity-time (PT) symmetric configurations with traceless Hamiltonians, here we have an unbalanced arrangement, where one of the two polarization components ($a_n$) is subject to a round trip dependent gain $g_n$ and the other one ($b_n$) to a detuning $\delta_n$ (phase mismatch). Clearly, for a parametric configuration of $\delta = 0$ and $g = 2\kappa$, the requirement for the emergence of a second order EP is met (in almost a PT symmetric configuration[15, 17, 18]). The quasi-adiabatic steering of the Hamiltonian in the gain-detuning parameter space can be achieved by establishing (discrete) temporal profiles for both the phase and gain as follows:

$$\delta_n = 2\kappa\rho x_\delta^{-1} \sin(\gamma 2\pi n/N), \qquad g_n = 2\kappa x_g^{-1}[1 - \rho\cos(\gamma 2\pi n/N)], \tag{2}$$

where $x_g$ and $x_\delta$ are gain and phase tuning parameters, and $\rho$ signifies the radius of a closed circular contour when $x_g = x_\delta$ (Fig. 1a). The choice of these parameters determines the shape of the parametric path. The direction of encirclement in equation (2) is dictated by the binary variable $\gamma$, where for $\gamma = 1$, a CW EP encirclement is executed, and $\gamma = -1$ leads to a CCW loop. The total number of round trips of the light pulse in the fiber ring, indicated by $N$ in equation (2), introduces a tunable degree of freedom to control the rate of change in gain and detuning after each cycle and thus the adiabaticity of the encirclement process. The onset of the dynamical encirclement lies in the PT unbroken phase to warrant a chiral response while avoiding unstable/undetectable modes (as a consequence of optical amplification/decay in the PT-symmetry broken phase[17, 18]).

Figure 1a illustrates the effect of the position of the parametric loop on the chiral mode conversion behavior. As can be seen, the directional response can still be observed even if the contour does not include an EP, provided that the winding is carried out sufficiently close to the EP and in a quasi-adiabatic manner[9]. Importantly, it shows that the chirality of mode conversion depends on the geometry of the parametric trajectories when the EP is not encircled. The degree of adiabaticity can be controlled by adjusting the coupling strength and the number of round trips $N$, which in turn affects the chirality of the process. These parameters can be modified at will so as to manipulate both the shape and relative position of the parametric loop with respect to the EP (Fig. 1b; see Supplementary Section 2). The quantity that measures the relative distribution of the eigenstate amplitudes over time is the eigenstate population $p(t) = [|c_1(t)|^2 - |c_2(t)|^2]/[|c_1(t)|^2 + |c_2(t)|^2]$. In this respect, $p$ refers to the relative weight of the eigenvector coefficients $c_1(t)$ and $c_2(t)$ associated with the state vector $|\psi(t)\rangle$ expanded in an eigenbasis, here the eigenvectors of the Hamiltonian at the onset of the parametric steering process, i.e., $|\psi(t)\rangle = c_1(t)|V_1\rangle + c_2(t)|V_2\rangle$. Evidently, this relative population ranges between $p \in [-1,1]$. The product of the relative eigenstate populations of the CW and CCW directions, $p_{cw} p_{ccw}$, provides a way to quantify the chirality of a parametric path, as depicted in Figs. 1a and b. In this way, a chirality function can be defined as $\chi = p_{cw} p_{ccw}$, that also takes values in the range $\chi \in [-1,1]$.

The contribution of the complex topology of the Riemann energy manifolds is brought to fore in Figs. 1c-e. For a loop centered at an EP (Fig. 1c) and starting from one eigenstate, an encirclement direction towards the amplified state (yellow plane) yields an adiabatic process and a state flip occurs. However, for the opposite direction, where the system initially tends to evolve with no amplification (green plane), the resulting non-adiabatic jump, a direct byproduct of the skewed vector space, returns the system to the initial eigenstate. These trajectories elucidate the fact that regardless of the initial state, the system always moves faithfully towards an eigenstate mandated by the direction of encirclement. Figure 1d shows that even by excluding the EP from this single-cycle parametric loop, the chirality of the process can still be retained depending on the proximity of the loop to the EP, provided that the Hamiltonian varies in a smooth fashion. Interestingly, to maintain the chiral state transfer, the field experiences an additional non-adiabatic jump likewise in both directions and for both initial states. On the other hand, shifting the trajectory further away from the EP (Fig. 1e) can deteriorate the chiral behavior.

A chiral mode conversion mechanism in polarization domain can be viewed as an omni-polarizer action (Fig. 2a). Depending on a CW or a CCW encirclement, any random input polarization state will arrive at one of the two diametrical (antipodal) points on the Poincaré sphere that represent the two mutually orthogonal eigenstates of the system. The operation principle of our proposed photonic emulator is schematically described in Fig. 2b (See Methods).

Of great importance in our experiments are the temporal and spatial stability of polarization states. A variable electrical polarization controller compensates the residual delay and ensures that the Mueller transfer function of the fiber loop is equal to the identity matrix. In addition, a soundproof box around the fiber loop shields the system from the effect of acoustic noise disturbances, and a judicious selection and configuration of the incorporated electro-optic and fiber devices establishes the adequate steadiness to perform the experiments. First, to confirm the stability of the setup, we measure the polarization state of a pulse during 50 round trips in the absence of intensity and phase tuning. Figure 3a shows that without polarization-dependent modulations the polarization remains the same, indicating that the system is indeed governed by the identity Mueller transfer matrix. A crucial part of this setup is the coupling between the two polarization components. Using a polarization controller, the constant coupling $\kappa$ is implemented via the induced retardation given that $\Gamma = 2\kappa$ (Supplementary Section 1). The effect of coupling was measured by examining the trajectory of the Stokes parameters of each pulse when three different initial states [$S_i^{(I)} = (-0.099, 0.24, -0.96)$, $S_i^{(II)} = (0.77, 0.55, -0.32)$, $S_i^{(III)} = (0.93, -0.14, 0.33)$] are probed on the Poincaré sphere with corresponding coupling levels $\kappa = 0.15, 0.25, 0.35$. Figures 3b, c compare the theoretically anticipated trajectories with those observed in the experiment. Note that the axis of rotation of the emerging circular trajectories remains constant under the exerted coupling strengths. Identifying this axis is important as its intersections with the Poincaré sphere reveal the two polarization eigenstates of the system.

Once stability is established, the parameter cycle is implemented by imposing the time-dependent phase and amplitude modulation [as stated in equation (2)]. Generally, for a 50-cycles loop (91.5 μs duration) a coupling of $\kappa \approx 0.7$ is required to ensure that the condition for quasi-adiabatic winding is satisfied (Fig. 1b). Pulses with different polarization states are launched into the system and their evolution in time is observed after each circulation in the ring. The Hamiltonian in the gain-detuning parameter space is steered in CCW direction along an elliptical trajectory ($x_g = 2.04, x_\delta = 4.2$) that excludes the EP albeit in its vicinity, for a pulse initiated in a polarization state with Stokes parameters $(-0.67, -0.17, -0.71)$, which is close to one of the eigenstates of the system $|V_1\rangle = (-0.78, 0.33, -0.53)$. The recorded evolution of the Stokes parameters shows to a great extent an adiabatic evolution that is intermittently disrupted by a non-adiabatic transition towards the other eigenstate $|V_2\rangle = (0.78, -0.33, 0.53)$, which is evidently the final state after 50 cycles. The trajectory of this evolution is displayed in Fig. 4a. On the other hand, traversing the parametric loop in the CW direction promotes two non-adiabatic jumps that force the final polarization state to return to the initial eigenstate with Stokes parameters $|V_1\rangle = (-0.78, 0.33, -0.53)$ (see Fig. 4b). As expected from theory, the two final states are indeed antipodal on the Poincaré sphere. To further verify that the system operates as a robust chiral mode convertor, we examine its response with another input polarization that happens to be closer to the complementary eigenstate $|V_2\rangle$. By initiating the process from this polarization state and letting the system evolve along the CW modulation pattern, the polarization of the pulse converges to a final state $|V_1\rangle$ after experiencing a non-adiabatic jump (Fig. 4c; see Supplementary Section 2). The dynamic effects related to modulating a non-Hermitian Hamiltonian in the vicinity of an EP can be best elucidated by projecting the evolving field $|\psi(t)\rangle$ onto the instantaneous eigenbasis at each time instant during the parametric steering process (see Supplementary Section 2). Generally, when the modulation is applied in a quasi-adiabatic manner, light tends to remain in the amplified eigenstate (with the largest imaginary part of the corresponding eigenvalue), until the followed eigenstate exchanges its amplification behavior with the other eigenstate (halfway through the loop in Fig. 4) and the evolution becomes unstable[31]. The onset of the ensuing non-adiabatic jump is, however, delayed. The skewed vector space close to the EP aids to prompt the non-adiabatic jump. This behavior is clearly displayed in Figs. 4d–f that show how the population of the eigenstates changes during the encircling process without encompassing an EP. The experimental results in Figs. 4g–i further confirm that the chiral state transfer reported here is caused by the topology and shape of the Riemann surfaces and the non-orthogonality of the modes, and not necessarily by the encirclement of an EP. The width (duration) and shape of the jumps depicted in Figs. 4d–i are highly influenced by how close the trajectories are to the EP, an aspect that makes them distinctly susceptible to noise and imperfections during calibration. This may lead to new and more complex mode conversion schemes by choosing different winding paths in parameter space and/or by introducing multiple exceptional points.

In conclusion, we have demonstrated a multifaceted platform to study the time-resolved dynamics of parametric steering of non-Hermitian Hamiltonians. Our experimental work clearly shows that chiral mode conversion is directly related to the landscape and topology of the Riemann surfaces and the skewedness of the vector space and not necessarily to the winding around an exceptional point, as previously thought. Our work differs from earlier demonstrations in which the analogy between time and propagation distance ($z$) is used to implement the necessary modulation profiles in space. In particular, the choice of polarization as a degree of freedom and our ability to map the evolution of light on the Poincaré sphere after each round trip, allow us to uniquely and with great precision expose the peculiarities of dynamical non-Hermitian systems in time. Our work may open new vistas in understanding the chiral and topological behaviors in non-Hermitian arrangements and can be used to develop a more comprehensive formalism for the adiabatic theorem in slowly varying non-Hermitian systems.

# Figure Legends

**Figure 1| Self-intersecting Riemann manifolds in the gain-phase parameter space and chiral mode conversion.**
**a**, A quantitative insight on the chirality of parametric steering of a Hamiltonian in the vicinity of a non-Hermitian singularity, performed via trajectories defined by $\delta_n = 2\kappa\rho x_\delta^{-1} \sin(\gamma 2\pi n/N)$ and $g_n = 2\kappa x_g^{-1}[1 - \rho\cos(\gamma 2\pi n/N)]$ with $N = 50$, $\kappa = 0.7$, $\rho = 1$, where $x_g$ and $x_\delta$ represent tunable gain and phase parameters. For a given set of parameters $(x_g, x_\delta)$, blue indicates a loop with a chiral response, whereas red denotes a non-chiral result. **b,** Adiabaticity as a prerequisite for a parametric loop sets a lower limit on the coupling strength between the two perpendicular polarization components and the number of round trips. The chirality map shown here, elucidates this limit for the case $x_g = x_\delta = 2.05$. **c,** Here, we numerically explain the chirality displayed by the dynamic variation of a non-Hermitian Hamiltonian in the gain-phase parameter space $[G = exp(g_n), \varphi = \delta_n]$ through the trajectories defined by equation (2) for $N = 300$, $\kappa = 0.3$, $\rho = 1$, $x_g = 1$, and $x_\delta = 1$ (red dots), when initiated from each polarization eigenstates on either of the Riemann surfaces (denoted as $s_i$ and marked by black square), for both clockwise CW and counterclockwise CCW encirclement helicities ($\gamma = \pm 1$, indicated by the black arrows underneath). The light's polarization evolves toward an eigenstate of the system, locked to the sense of rotation, no matter what the initial state is. The final state is indicated by $s_f$ and is represented by a white circle. **d-e,** same as **c,** but when the EP is not included in the parametric loops. In **d,** the EP still lies in the vicinity of the loop while in **e** it is positioned far away. Red dots in **c-e,** numerically obtained by solving the difference equation (S3) and the consecutive projection of the dynamical evolution onto the Riemann surfaces via $[Re(\lambda_1(t))|c_1(t)|^2 + Re(\lambda_2(t))|c_2(t)|^2]/[|c_1(t)|^2 + |c_2(t)|^2]$, represent the evolution of the eigenvalues of the system during the parametric steering process (See Supplementary Section 2). The underlying Riemann surfaces are associated with the real part of the eigenvalues of the corresponding Hamiltonian of the system with the green and yellow color indicating (relative) loss and gain, respectively.

**Figure 2| Operation principle of the devised fiber-based photonic emulator. a**, Conceptual illustration of the chiral mode conversion in the polarization domain as a result of dynamic steering of a non-Hermitian Hamiltonian around its singularity in parameter space. The polarization is funneled into one of the orthogonal eigenstates of the system and is solely determined by the helicity of the parametric loop. **b,** Experimental setup: the seed pulse module is composed of a laser diode and an intensity modulator (IM) for generating pulses, an erbium-doped fiber amplifier (EDFA) for pulse amplification, an acousto-optic modulator (AOM) and a tunable bandpass filter for removing the excess noise introduced by the EDFA and a polarization controller (PC) for adjusting the initial polarization state. A 300 m single mode fiber (SMF) sets the pulse round trip time and the AOM acts as a gate, controlling the number of round trips (Supplementary Section 3). The variable polarization controller (VPC) calibrates the setup and couples the polarization components. The desired gain-phase variation of the two polarizations is realized by using a polarization beam splitter (PBS), an intensity modulator, a phase modulator (PM), a delay line (DL), a variable optical attenuator (VOA) and a polarization beam combiner (PBC). The fiber Bragg grating (FBG) removes the pilot wavelength after the EDFA and purifies the signal from any amplified spontaneous emission noise (ASE) from the EDFA. At the end of each round trip, half of the pulse is routed to a monitoring module, equipped with an InGaAs photo detector (PD), a PicoScope and a polarimeter, while the other half is launched back into the fiber loop for recirculation at predetermined times.

**Figure 3| Polarization stability and coupling between the polarization components of a light pulse. a**, Demonstrating the stability of the light pulse's polarization state over the probed time span of the experiment (over 50 round trips of the pulse in the fiber loop: 91.5 $\mu s$), when the polarization components are uncoupled and the intensity and phase modulators are deactivated. The polarization of the light at the end of each round trip (shown by red circles) has remained localized on the Poincaré sphere. This step is critical given that this fiber-based emulator is interferometric in nature and is thus susceptible to noise sources. LCP: Left-handed Circular Polarization, RCP: Right-handed Circular Polarization, LHP: Linear Horizontal Polarization, LVP: Linear Vertical Polarization, L +45P: Linear +45 Polarization, L -45 Polarization: Linear -45 Polarization **b-c,** Stokes vector representation of the polarization evolution of a light pulse when its polarization components are coupled to each other at a fixed strength. In this case, irrespective of the coupling level and initial conditions, the polarization state should move on circular trajectories on the Poincaré sphere that share the same axis of rotation, as also confirmed in the experiment. The end points of this axis represent the polarization eigenstates of the system. Theoretical simulations in **b** are in good agreement with the experimental results displayed in **c.**

**Figure 4| Experimental observation of chiral mode conversion through a parametric steering of the Hamiltonian along a trajectory that lies in the proximity of an EP. a**, The system is excited with a light pulse that is randomly polarized (i.e. in an arbitrary mixture of the polarization eigenstates, denoted as $S_i$ and displayed by the green diamond), which happens to be close to the first eigenstate $|V_1\rangle$ of the Hamiltonian. This state is forced to traverse an EP-excluding parametric loop, albeit in its vicinity, in the gain-phase parameter space, characterized by the modulation patterns $\delta_n = 2\kappa\rho x_\delta^{-1}\sin(\gamma 2\pi n/N)$, $g_n = 2\kappa x_g^{-1}[1-\rho\cos(\gamma 2\pi n/N)]$ with $x_g = 2.04$ and $x_\delta = 4.2$. This loop is carried out in the CCW direction ($\gamma = -1$), through 50 round trips in the fiber ring when the coupling strength between the orthogonal polarizations is $\kappa \approx 0.7$. The experimental results demonstrate a faithful evolution of the polarization toward the second eigenstate $|V_2\rangle$ of the system (represented by $S_c$ and displayed by a red star). The final state $S_f$ observed in the experiment is marked by a yellow circle. **b**, Starting close to $|V_1\rangle$ but steering the system with the opposite helicity ($\gamma = 1$; CW), the polarization robustly converges to the eigenstate $|V_1\rangle$ that is located at a diametrical point on the Poincaré sphere. **c**, By keeping the helicity of encirclement as in **b** but with different initial conditions ($|V_2\rangle$), the end result remains the same: a faithful conversion towards $|V_1\rangle$–an aspect that substantiates the chirality of the process in the polarization domain. LCP: Left-handed Circular Polarization, RCP: Right-handed Circular Polarization, LHP: Linear Horizontal Polarization, LVP: Linear Vertical Polarization, L +45P: Linear +45 Polarization, L -45 Polarization: Linear -45 Polarization. **d–f**, Numerical evolution of the polarization state of the system according to equation (S3) when projected on the instantaneous eigenstates and **g–i**, experimentally measured polarization state projected onto the same eigenstates as in **d-f**. Error bars represent the uncertainty in the extraction of the Stokes parameters from the data collected by the polarimeter. The experimental trajectory in the gain-detuning parameter space is displayed as inset in panel **i**. Solid curves are obtained by numerically fitting the experimental data. Discrepancies between theory and experiment stem mostly from the uncertainty of the instantaneous eigenstates in the experiment (see Supplementary Section 2)

**Methods**

In our setup, shown in Fig. 2b, a 330 ns laser pulse at a wavelength of 1547 nm (generated by a semiconductor laser and an intensity modulator) featuring an arbitrary state of polarization is injected in the main loop through a 3 dB coupler. The single mode fiber in use is not polarization maintaining. Here, we refer to the two orthogonal polarization states as $P_1$ and $P_2$. A two-dimensional parameter space is formed through the gain and detuning (phase difference between $P_1$ and $P_2$). A polarization beam splitter directs the vertical and horizontal field components into two separate branches where they independently yet simultaneously experience gain and phase modulation, imposed by electro-optic intensity and phase modulators, respectively, with a temporal dependence as described in equation (2). The branches are equalized in terms of their length and insertion loss by deploying a variable optical attenuator (VOA) and a delay line (DL). The two beams are then combined by means of a polarization beam combiner. Finally, a constant coupling is provided by a polarization controller that mixes the polarization states. At the end of each round trip, half of the pulse couples out of the fiber ring for monitoring purposes and the rest remains in the loop to repeat the process [with the modulation patterns stated in equation (2)], until the desired number of pulse circulations is attained. The part of the pulse that exits the loop in each cycle, provides a unique opportunity for an in-situ and real time monitoring of the light's polarization evolution during the encirclement process. In this setup, the degree of adiabaticity can be controlled with the aid of two acousto-optic modulators (AOMs) that gate pulse circulation and therefore set the desired encirclement time. To prevent pulses from fading, an erbium-doped fiber amplifier (EDFA) is inserted in the loop, in order to compensate the insertion loss introduced by the electro-optics and fiber components (Supplementary Sections 3). In the ring, a 300 m long single mode fiber (SMF) ensures the presence of only one pulse during the total encirclement process (91.5 μs). Each round trip lasts approximately 1.83 μs. The polarimeter used in this study (NOVOPTEL PM1000) has a sampling rate of 100 MHz which allows to properly characterize pulses with a duration of 330 ns.

**Acknowledgements**
We gratefully acknowledge the financial support from the Air Force Office of Scientific Research (Multidisciplinary University Research Initiative (MURI) Award on Novel light-matter interactions in topologically non-trivial Weyl semimetal structures and systems: FA9550-20-1-0322, MURI Award on Programmable systems with non-Hermitian quantum dynamics: FA9550-21-1-0202), DARPA (D18AP00058), the Office of Naval Research (N00014-19-1-2052, N00014-20-1-2522, MURI Award on Classical entanglement in structured optical fields: N00014-20-1-2789), the Army Research Office (W911NF-17-1-0481), the National Science Foundation (DMR-1420620, EECS-1711230, ECCS CBET 1805200, ECCS 2000538, ECCS 2011171), the W. M. Keck Foundation, the US–Israel Binational Science Foundation (BSF; 2016381), the MPS Simons collaboration (Simons grant 733682), the US Air Force Research Laboratory (FA86511820019), the Austrian Science Fund (FWF, P32300 WAVELAND) and European Commission grant MSCA-RISE 691209. G.L-G. acknowledges support from Consejo Nacional de Ciencia y Tecnologia (CONACyT). We thank A. Turchanin and U. Peschel from Friedrich Schiller University Jena for useful discussions and feedback.

**Author contributions**
D.N.C. and M.K. conceived the idea. H.N., G.L-G. and H.E.L-A. designed and performed the experiments in consultation with other team members. H. N., A.U. H. and A.S. performed the analysis with helps from other members. H. N., M.K. and D.N.C. wrote the manuscript with help from all the authors.

**Competing interests**
The authors declare no competing interests.

**Supplementary Information** is available for this paper.

**Additional information**
**Reprints and permissions information** is available at www.nature.com/reprints.

**Correspondence and requests for materials** should be addressed to
khajavik@usc.edu

**Data availability**

All data that support the findings of this study are available within the paper and the Supplementary Information and are available from the corresponding author upon request.

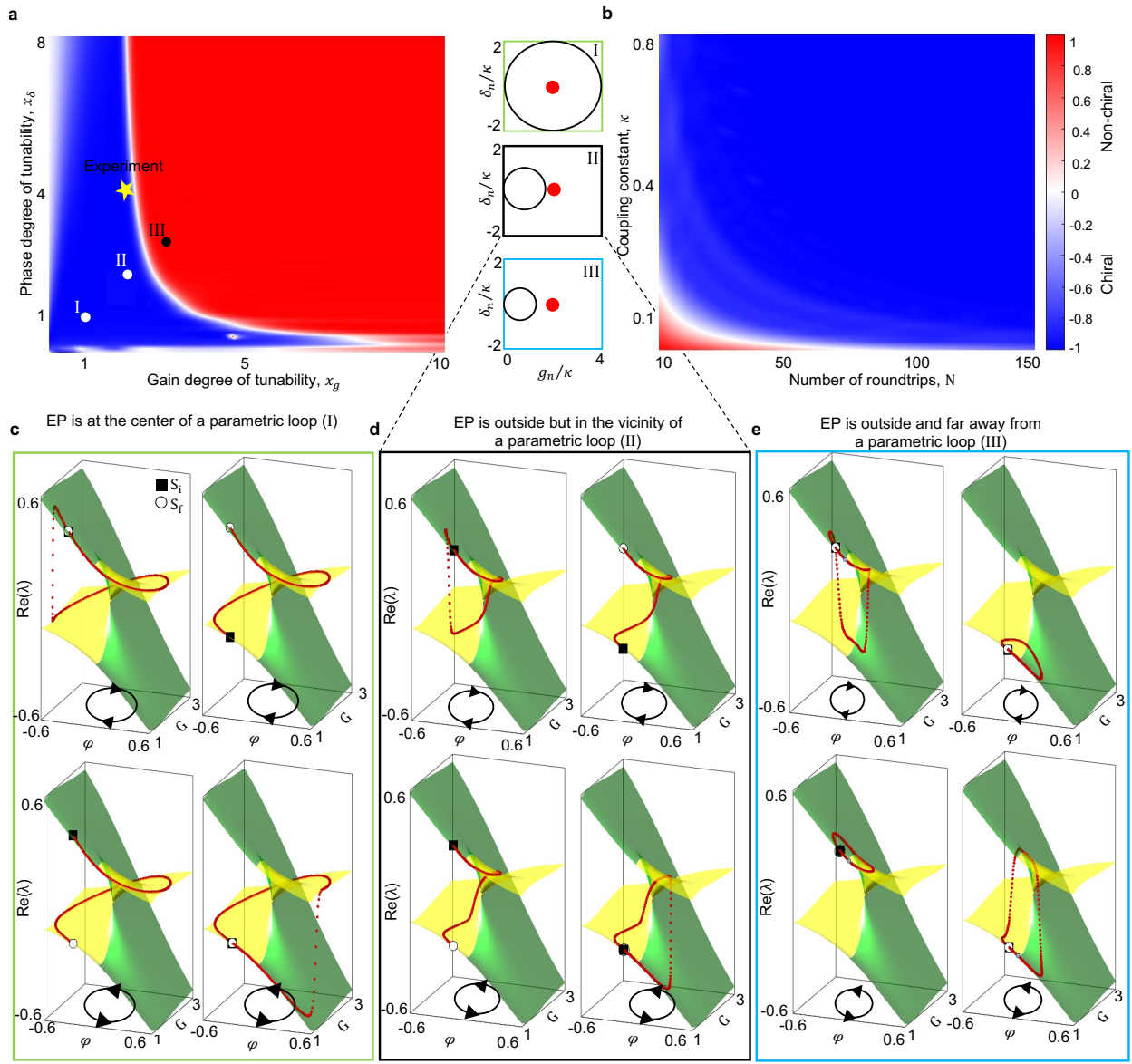

**Figure 1|**

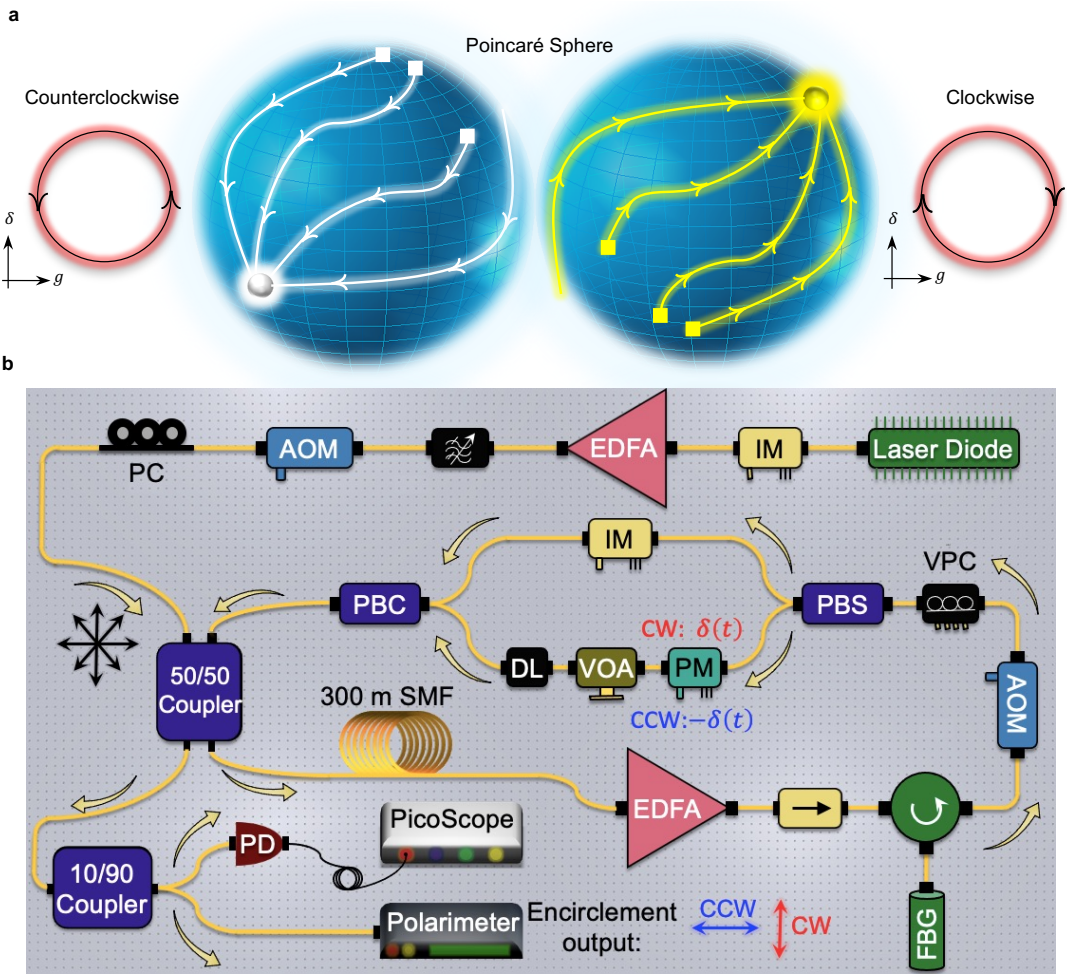

**Figure 2|**

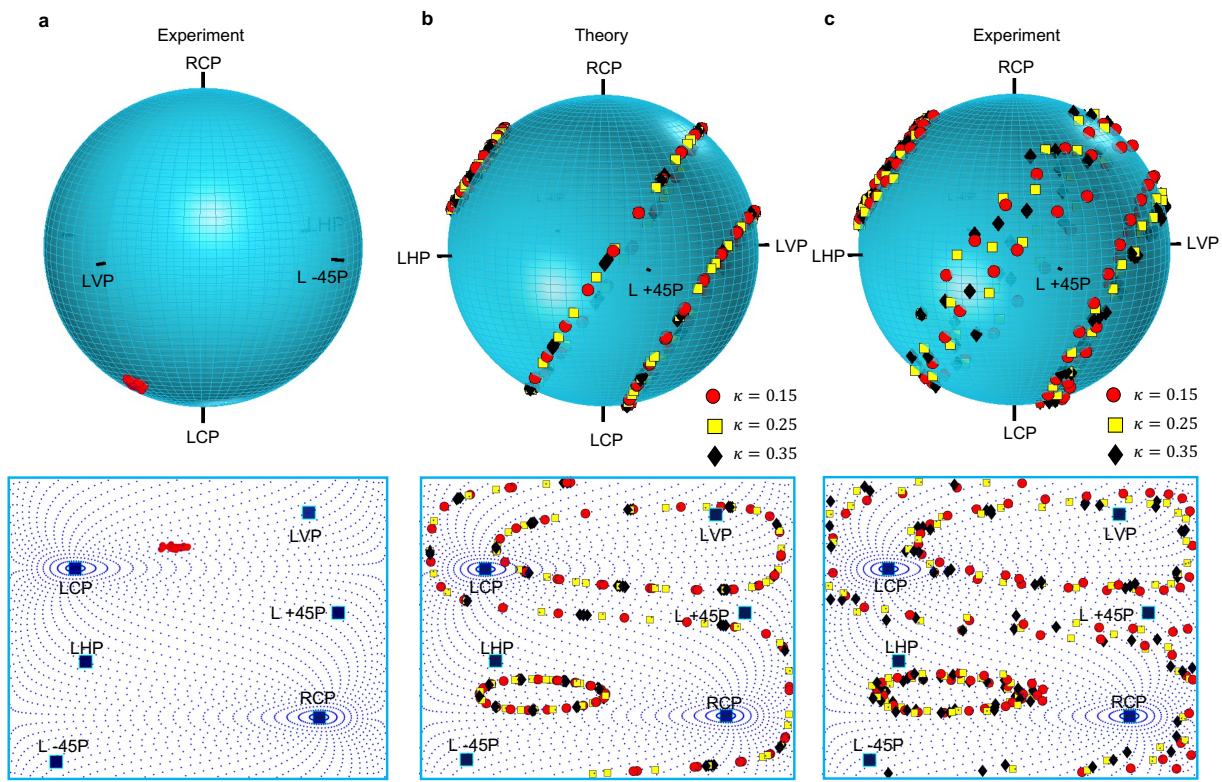

**Figure 3|**

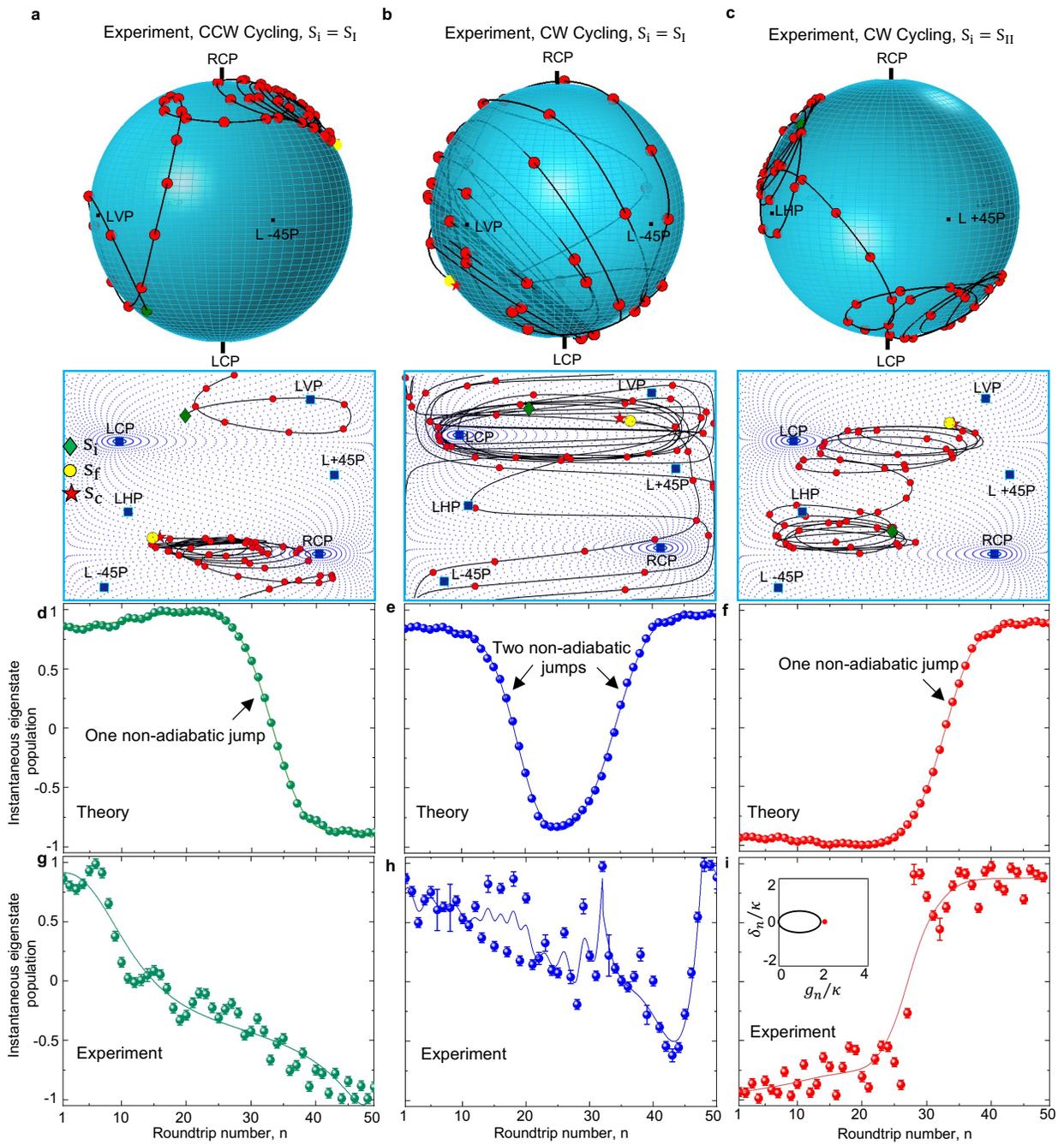

**Figure 4|**